\documentclass[%
 aip,
 sd,%
 amsmath,amssymb,
 preprint,%
]{revtex4-1}

\usepackage{array}
\usepackage{graphicx}% Include figure files
\usepackage{dcolumn}% Align table columns on decimal point
\usepackage{bm}% bold math
\usepackage{mathrsfs,amsmath} 
\usepackage{mathtools}

\usepackage{enumerate}
\usepackage[]{subfigure}
\begin{document}

%\preprint{AIP/123-QED}

\title{Calculation of x-ray scattering patterns from nanocrystals at high x-ray intensity}

\author{Malik Muhammad Abdullah}
\affiliation{ Center for Free-Electron Laser Science, DESY, Notkestrasse 85, 22607 Hamburg, Germany }
\affiliation{ The Hamburg Centre for Ultrafast Imaging, Luruper Chaussee 149, 22761 Hamburg, Germany }
\affiliation{ Department of Physics, University of Hamburg, Jungiusstrasse 9, 20355 Hamburg, Germany }

\author{Zoltan Jurek}%
\affiliation{ Center for Free-Electron Laser Science, DESY, Notkestrasse 85, 22607 Hamburg, Germany }
\affiliation{ The Hamburg Centre for Ultrafast Imaging, Luruper Chaussee 149, 22761 Hamburg, Germany }

\author{Sang-Kil Son}
\affiliation{ Center for Free-Electron Laser Science, DESY, Notkestrasse 85, 22607 Hamburg, Germany }
\affiliation{ The Hamburg Centre for Ultrafast Imaging, Luruper Chaussee 149, 22761 Hamburg, Germany }

\author{Robin Santra}
\affiliation{ Center for Free-Electron Laser Science, DESY, Notkestrasse 85, 22607 Hamburg, Germany }
\affiliation{ The Hamburg Centre for Ultrafast Imaging, Luruper Chaussee 149, 22761 Hamburg, Germany }
\affiliation{ Department of Physics, University of Hamburg, Jungiusstrasse 9, 20355 Hamburg, Germany }

\date{\today}

\begin{abstract}
We present a generalized method to describe the x-ray scattering intensity of the Bragg spots in a 
diffraction pattern from nanocrystals exposed to intense x-ray pulses. 
Our method involves the subdivision of a crystal into smaller units. 
In order to calculate the dynamics within every unit 
we employ a Monte-Carlo (MC)-molecular dynamics (MD)-ab-initio hybrid 
framework using real space periodic boundary conditions. 
By combining all the units we simulate the diffraction pattern of a crystal larger than the transverse x-ray beam profile, a situation
commonly encountered in femtosecond nanocrystallography experiments with focused x-ray free-electron laser radiation. 
Radiation damage is not spatially uniform and depends on the fluence associated with each specific region inside the crystal. To investigate the effects of 
uniform and non-uniform fluence distribution we have used two different spatial beam profiles, 
gaussian and flattop.
\end{abstract}

\maketitle

\section{\label{sec:level1}Introduction}
With the advent of x-ray free electron laser (XFEL) sources~\cite{norah}, studies of structural determination of 
biomolecules~\cite{chapman2011,Barty2012,boutet2012,Redecke2013} have gained a new boost. XFELs provide intense radiation of a wavelength 
comparable to atomic scales. The characteristics of XFEL radiation and associated sample environments have triggered the development of 
new data collection methods, such as serial femtosecond crystallography~\cite{chapman2015} (SFX). The ultimate goal and dream is to 
perform atomic resolution single particle imaging~\cite{Neutze2000,michael,haureige2005,chapman2010,aquila2015}. Sample damage by 
x-rays and low signal to noise ratio at high photon momentum transfer limit the resolution of structural studies on non-repetitive structures such as individual biomolecules 
or cells~\cite{haureige2007,Neutze2000}. Therefore at high resolution SFX is currently still a better option to use. XFELs deliver intense 
femtosecond pulses that promise to yield high-resolution diffraction data of nanocrystals ($\sim$200\,nm to 2 $\rm{\mu}$m in size) before 
the destruction of the sample by radiation damage~\cite{neutze2014,caleman2014}. In SFX, a complete data-set can be obtained by exposing 
thousands of randomly oriented, individual crystals of proteins to the x-ray beam.

For imaging proteins and viruses at atomic resolution one calls for high intensity and short x-ray 
pulses~\cite{zoltan2004,barty2008,Neutze2000,Garman2010,boutet2010,caleman2011}.
The shortcoming of high intensities is the rapid ionization of the atoms on the few femtosecond timescale, which affects the structure of 
the system. This radiation induced damage changes the atomic form factors~\cite{quiney,sangyoulinda} and may induce significant atomic displacement 
on longer times. Finally radiation damage changes the scattering pattern.
For a comprehensive theoretical study of signal formation in an SFX experiment one needs to simulate (i) the radiation induced dynamics of the sample 
and (ii) pattern formation based on the dynamics.
During the past decade, several models have been developed for studying the time evolution of small and large samples irradiated by XFEL pulses~\cite{Scott2001,
bergh2004,hauriege2004, zoltan2004-2,beata2006, saalmann2009,caleman2011-2,haureige2012,Fang2012}.  We use XMDYN~\cite{zoltan2014,tachibana2015}, 
a Monte-Carlo molecular-dynamics based code developed by the authors.
In the theoretical study presented here we consider a micron-size crystal in a 100\,nm focus beam, a scenario where a nanocrystalline sample experiences
fluences as high as to be used in single particle imaging experiments. As a consequence, the x-ray 
fluence is non-uniform throughout the sample. This may also have its imprint in the scattering pattern.
The bottleneck one faces is that it is computationally not feasible to simulate a system with realistic size using tools which are capable to follow the dynamics of 
each atom, required for imaging studies. 
Therefore, we present an approach that involves the divison of a crystal into smaller units (super-cells) and the calculation of their dynamics individually using periodic boundary 
conditions (PBC). In order to investigate the effect of inhomogeneous spatial fluence distribution, the super-cells are subjected to different fluences. 
Then we combine all the super-cells to form a nano-crystal and construct the scattering pattern under the influence of uniform (within the irradiated part of the sample) 
and non-uniform spatial beam profiles. We study and compare these two scenarios. 

\section{\label{sec:level1}Methodology}
\subsection{\label{sec:level2}Radiation damage simulation}

XMDYN~\cite{xatom-xmdyn,zoltan2014,tachibana2015} has been originally developed for modeling finite-size systems irradiated by an XFEL pulse. It unites
a Monte-Carlo description of ionizations with a classical molecular-dynamical treatment of particle dynamics. XMDYN keeps 
track of the configuration of the bound electrons in neutral atoms and atomic ions. These configurations change dynamically because of 
different atomic processes like inner and outer-shell photoionization, Auger and fluorescence decay and collisional (secondary) ionization.  

In order to treat x-ray-atom interactions, XMDYN uses the XATOM~\cite{sangyoulinda,sangkil2012} toolkit, which is an ab-initio framework based on non-relativistic quantum 
electrodynamics and perturbation theory. XATOM provides rates and cross-sections of x-ray-induced processes such as photoionization, Auger decay and 
x-ray fluoresence. XMDYN employs XATOM data, keeps track of all the ionization events along with the electron configuration of each atom, calculates impact ionization and recombination and follows the trajectories of all the ionized 
electrons and atoms solving the classical equations of motion numerically. 
%XMDYN includes microscopic processes and plasma properties that appears as an outcome like nanoplasma formation, screening, thermalization of electrons through collisions 
%and thermal emission~\cite{tachibana2015}. Moreover the bonds are not considered as the fluence is considerably high which leads to immediate ionization of the atoms~\cite{zoltan2014}.   
The framework based on these microscopic processes can describe complex many-body phenomena in ionized systems such as nanoplasma formation, charge screening, thermalization of electrons through collisions 
and thermal emission~\cite{tachibana2015}. In the current study chemical bonds between carbon atoms are not considered. This is a good approximation when the fluence is high enough to cause severe ionization in the system early in the pulse. 
The immediate ionization of the atoms leads to fast bond breaking that allows their exclusion in simulations~\cite{zoltan2014}.

\subsection{\label{sec:supercell}Super-cell approach}

The dimensions of the interaction volume are defined by the intersection of the x-ray beam and the crystal, therefore its dimensions are determined by 
the focal area ($\sim 100\times100\,\rm{nm}^2$) and the thickness of the crystal 
along the beam propagation direction ($\rm{\mu} m$). The number of atoms within this volume 
is of the order of 10$^{9}$. This number is formidably large: it is not feasible to simulate the whole system by a single XMDYN run.
In order to overcome this barrier we propose the procedure of dividing the whole crystal into smaller units. These super-cells may contain several 
crystallographic unit cells. We follow the dynamics within each super-cell driven by the local fluence (assumed to be uniform throughout the super-cell) individually.
For this purpose we have developed an extension to XMDYN that applies PBC~\cite{ewald,ewald2} to a super-cell, accounting also for the effect of the environment surrounding it.
 
Within the concept of PBC, a hypothetic crystal is constructed as a periodic extension of a selected super-cell. The total Coulomb interaction 
energy for a super-cell includes all the interactions within the given cell as well as pair interactions when 
one particle is in the cell while the other is in a periodic image within the super-cell based hypothetic lattice (PBC-crystal). Formally:
\begin{eqnarray}
\label{eq:sum}
E = \frac{1}{4\pi\varepsilon_{0}} \frac{1}{2} \sum_{\textbf{n}} \sum_{i=1}^{N} \sum_{j=1}^{N}{'} \frac{q_{i} q_{j}}{|\textbf{r}_{ij} + \textbf{n}L|} 
\end{eqnarray}
where $N$ represents the total number of particles in the super-cell, $q_i$ is the charge of the $i$th particle, $\varepsilon_{0}$ is the dielectric 
constant, $L$ represents the dimension of the cell (here assumed to be a cube), 
$\textbf{n}L = n_{1} \textbf{c}_{1} + n_{2} \textbf{c}_{2} + n_{3} \textbf{c}_{3} $, where  
$\textbf{c}_{1}, \textbf{c}_{2}, \textbf{c}_{3}$ represent basis vectors of the PBC-crystal, 
and $n_{1},n_{2},n_{3}$ are integers indexing the periodic images. Hence, ${|\textbf{r}_{ij} + \textbf{n}L|}$ is the distance between the $i$th particle in the 
central super-cell ($\textbf{n}=\textbf{0}$) and $j$th particle in the super-cell indexed by $\textbf{n}$. The symbol $'$ represents the exclusion of the term $j=i$ if
and only if $\textbf{n} = \textbf{0}$. The summation in Eq.~\ref{eq:sum} is not only computationally very expensive because of the formally infinite sum  
but is also conditionally convergent which states that the result depends upon the order of summation.
To overcome this problem we follow a route used often in the literature for spatially periodic systems, the method of minimum image convention~\cite{metropolis}.
According to the convention: 
(i) when evaluating Eq.~\ref{eq:sum}, we do not use the same super-cell division of the PBC crystal for all particles, but we always shift the boundaries so that the 
selected particle appears in the center ; 
(ii) we consider only $\textbf{n} = \textbf{0}$ terms. 
The former choice ensures that no jump happens in the potential energy when a particle crosses a super-cell boundary and therefore 'jumps' in the evaluation from one 
border of the cell to the opposite. The latter is a minimum choice considering interactions between a selected particle with the closest copy of the others only. 
Finally one can assemble the entire real crystal from the individually simulated
super-cells to model the whole dynamics. While in this way modeling becomes
feasible even without the need of super-computers, we should also note a
shortcoming of the approach: we do not allow particle transport, in particular
electron transport between the super-cells. For biologically relevant light
elements Auger and secondary electrons have energies $E_{kin} \lesssim 300\,\rm{eV}$, which
yields a short mean free path in a dense environment. Therefore, such electrons may
travel only to neighboring super-cells experiencing similar fluences
during the irradiation, so that the effect of net transport may be
negligible. On the other hand photoelectrons have an energy almost as high as the photon energy. Hence, they are fast and have a long mean free
path: they can leave super-cells located at high fluences regions and can affect
super-cells at larger distances experiencing lower fluences. We will overcome this shortcoming of the model in the future.

\subsection{\label{sec:level2}Scattering intensity}

Although during a single shot experiment the sample may undergo significant
changes, the scattering patterns are static: they accumulate diffracted signal
over the whole pulse. Further, the signal may contain an imprint of a spatially
non-uniform intensity profile. Formally, the scattering intensity at a specific
reflection described by reciprocal vector $\textbf{Q}$, including the integration over
time and the subdivision of the crystal volume into super-cells according to the approach introduced in Section \ref{sec:supercell}, 
reads:
 
\begin{eqnarray}
\label{eq:formal}
 \frac{dI(\textbf{Q},\mathcal{F},\omega)}{d\Omega} = C(\Omega)\int_{-\infty}^{\infty}dt \, g(t)\sum_{I,r} P_{I,r}
(\mathcal{F},\omega,t)\nonumber \\
\left|{ \sum_{\mu} \sqrt{\mathcal{F}_{\mu}} \, e^{\textit{i}\textbf{Q}\cdot \textbf{R}_{\mu}} \, \sum_{X}\sum_{j=1}^{N_X} \,\, \textit{f}_{X,I_{X,j}^{\mu}}(\textbf{Q},\omega) \, 
e^{\textit{i}\textbf{Q}\cdot\textbf{r}_{X,j}^{\mu}} }\right|^2 
\end{eqnarray}
In this equation, $\textbf{Q}$ is the momentum transfer, $\mathcal{F}=\{\mathcal{F}_{\mu}\}$ is the 
x-ray fluence distribution throughout the crystal, the index $\rm{\mu}$ runs over all super-cells and $\omega$ 
is the photon energy.  $C(\Omega)$ is a factor depending the polarization of the x-ray pulse, $g(t)$ represents 
the normalized temporal envelope. $\textit{f}_{X,I_{X,j}^{\mu}}$ is the atomic form factor 
of the $j$th atom of species $X$ in the $\rm{\mu} th$ super-cell, $I_{X,j}^{\mu}$ is the associated electronic configuration, $I=\{I_{X,j}^{\mu}\}$ 
denotes a global electronic configuration, $\textbf{r}_{X,j}^{\mu}$ 
represents the position vector of the $j$th atom of species $X$ in the $\rm{\mu th}$ super-cell, and $r=\{\textbf{r}_{X,j}^{\mu}\}$ indicates 
the set of all atomic positions. $N_{X}$ represents the total number of atoms for species $X$ within a super-cell. $P_{I,r}$ represents the probability distribution 
of electronic configuration $I$ and atomic positions $r$, and $\textbf{R}_{\mu}$ represents the position of the $\rm{\mu th}$ 
super-cell. The atomic form factor 
\begin{eqnarray}
\label{eq:dispersion}
\textit{f}_{X,I_{j}^{X}}(\textbf{Q},\omega) = \textit{f}_{X,I_{X,j}^{\mu}}^{\,0}(\textbf{Q}) +
\textit{f}_{X,I_{X,j}^{\mu}}^{\,'} (\omega) + \textit{i} \, \textit{f}_{X,I_{X,j}^{\mu}}^{\,''}(\omega)
\end{eqnarray}
includes the dispersion corrections $\textit{f}_{X,I_{X,j}^{\mu}}^{\,'} (\omega)$ and $\textit{i} \, \textit{f}_{X,I_{X,j}^{\mu}}^{\,''}(\omega)$. 
This dispersion correction can be neglected when the applied photon energy is high above the ionization edges, which is fulfilled in our 
study. Note that the summation over $\sqrt{\mathcal{F}_{\mu}}$ appears inside the modulus square in Eq.~\ref{eq:formal}. The scattering 
amplitude from the $\rm{\mu th}$ super-cell is proportional to the x-ray field amplitude 
($\propto\sqrt{\mathcal{F}_{\mu}}$) in that super-cell. A key assumption when performing the coherent sum in Eq.~\ref{eq:formal} is that 
the entire crystal is illuminated coherently, a condition that is fulfilled considering realistic XFEL beam parameters and crystal sizes. 

%which is the reason why the scattering amplitudes from all super-cells are added coherently in Eq. (3). 
\subsection{\label{sec:consider}XSINC: Scattering pattern simulation}
In order to construct the scattering pattern,  Eq.~\ref{eq:formal} cannot be used
directly as the $P_I$ and $r$ configuration space is too large. However, by
calculating realizations of super-cell dynamics with XMDYN, a
Monte-Carlo sampling of the distribution $P_{I,r}(\mathcal{F},\omega,t)$ represented in Eq.~\ref{eq:formal} becomes feasible. To construct the time evolution of the crystal through
global configurations and to calculate patterns, we used the following strategy, implemented in the code XSINC (x-ray scattering in nano-crystals). 

\begin{figure*}
\includegraphics[width=12.2cm]{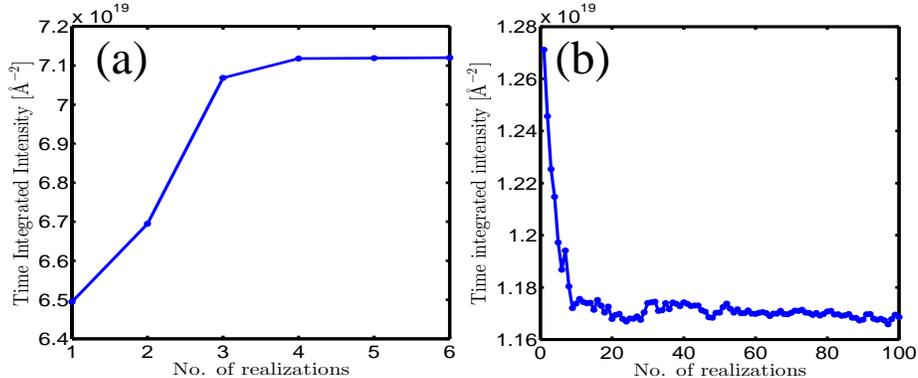}
\caption{ Convergence of time integrated peak intensity for the reflection (1 1 1) as a function of the number of realizations per fluence point: 
(a) for the gaussian case and (b) the flattop case. For the gaussian case, 350 different fluences points are used to calculate the time integrated intensity.}
\label{fig:convergence-realiz}
\end{figure*}

We discretize the fluence space and calculate many super-cell trajectories for each fluence value with XMDYN. 
XSINC selects randomly a trajectory for each super-cell within the crystal (a local realization), so that the corresponding fluence values are matching the best. 
These trajectories describe the local time evolution of the super-cells and together they form a global realization of the crystal. 
Then, taking into account the spatial and temporal pulse profiles, XSINC calculates the scattering amplitudes and intensities for the global 
configuration at different times based on the corresponding snapshots.
Finally, the incoherent sum of these patterns corresponds to a time integrated pattern measured at in a single-shot experiment. 
In our calculation we perform a dense sampling of the fluence space. As a consequence, two neighboring super-cells 
experience very similar fluence. Therefore it is a good approximation to take into account the direct effect of the neighboring cells by applying periodic 
boundary conditions and this construction leads to a realistic global trajectory. In the scheme above several 
parameters are convergence parameters of the method (Table~\ref{table:converg_param}). Results are considered 
converged when characteristic properties of the Bragg peaks, such as the width and height of the intensity distribution 
in reciprocal space, converge during monotonic increase (or decrease) of the parameter. As an example Figures~\ref{fig:convergence-realiz}(a) and~\ref{fig:convergence-realiz}(b)
illustrates the convergence of the time integrated peak intensity as a function of the number of local (super-cell) realizations per fluence point for the reflection (1 1 1) 
for the gaussian and flattop spatial profile cases. We note that convergence implicitly depends on the total number of different realizations 
used to build a global realization. Therefore in the gaussian case, where 350 different fluence points are used, convergence starts at a much smaller value.

\begin{table*}
\begin{tabular}{|p{7cm}|p{4cm}|p{4cm}|}
\hline
\textbf{Convergence Parameters}& \textbf{Gaussian Case} & \textbf{Flattop Case}\\
\hline
Number of crystallographic unit cells in a super-cell    & $5 \times 5 \times 5 $ & $5 \times 5 \times 5 $\\
Number of fluence points   & 350    & 1\\
Number of local realizations (XMDYN trajectories) per fluence point&   5  & 150\\
Number of assembled global realizations &10 & 10\\
Depth of the crystal in beam propagation direction& 1\,$\times$\,Thickness of the super-cell lattice constant  & 1\,$\times$\,Thickness of the super-cell lattice constant\\
Number of snapshots& 28  & 28   \\
\hline
\end{tabular}
\caption{Convergence parameters for calculating scattering intensity with XSINC and their values in the current study.}
\label{table:converg_param} 
\end{table*}

\section{\label{sec:level1}Results and Discussion}

\subsection{\label{sec:Setup}Simulation setup}
\begin{figure}
\includegraphics[width=7.0cm,height=6.0cm]{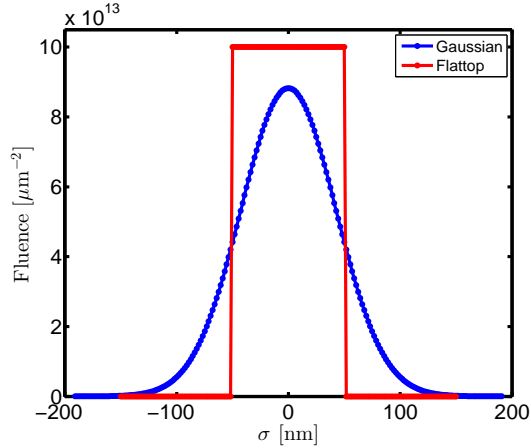}
\caption{ Radial fluence distributions in the current study: gaussian profile (spatially non-uniform case) and flattop profile (uniform within the irradiated part of the crystal).
The focal size is 100\,nm in both cases and the pulse energy is also considered to be same.
}
\label{fig:fluence.dist}
\end{figure}

In our investigations we consider a diamond cube of a size of 1\,$\rm{\mu}$m. We investigate the cases of flattop 
and gaussian beam profiles (Fig.~\ref{fig:fluence.dist}). Other parameters of the pulses are the same in both cases: 
photon energy is 10\,keV, total number of photons per x-ray pulse is 1$\times$10$^{12}$,
the temporal pulse envelope is gaussian with a duration of 10\,fs FWHM, focus size is 100$\times$100\,$\rm{nm^2}$ FWHM. 
The size of the diamond unit cell is $a=b=c=$ 3.57\,$\rm{\AA}$ containing 8 carbon atoms. 
The parameter choices listed in Table~\ref{table:converg_param} yield converged results.

\subsection{\label{sec:Damage}Radiation damage}

\begin{figure*}
\includegraphics[width=12.2cm]{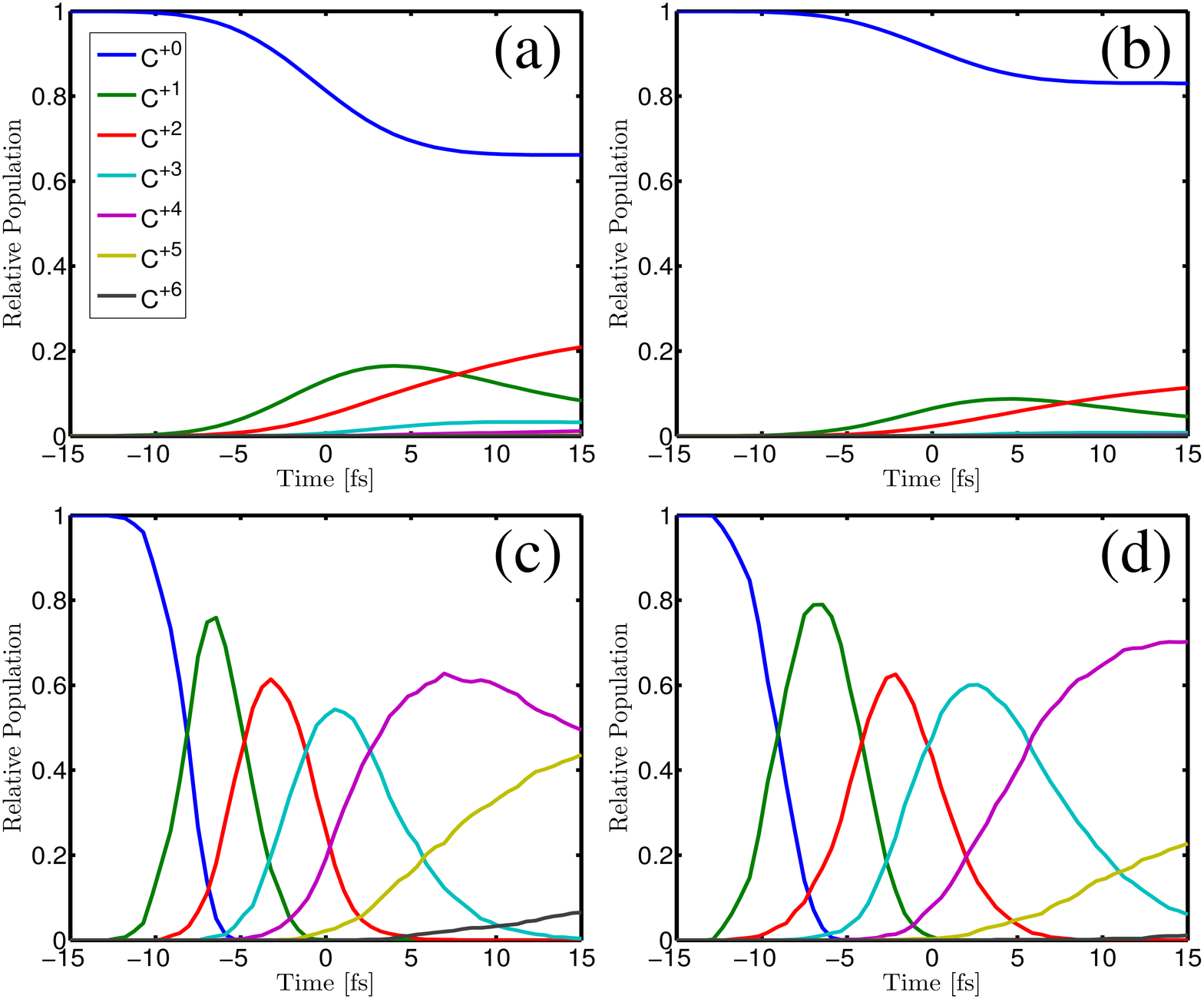}
\caption{Ionization dynamics of carbon atoms at different fluences:
time dependent charge state populations of isolated carbon atoms calculated with XATOM for (a) $\mathcal{F}_\text{high} = 1\times10^{14}\,\rm{\mu} m^{-2}$ and 
(b) $\mathcal{F}_\text{mid} = 4.5\times10^{13}\,\rm{\mu} m^{-2}$. Similarly, time dependent charge state populations of carbon atoms in diamond calculated with 
XMDYN for (c) $\mathcal{F}_\text{high}$ and (d) $\mathcal{F}_\text{mid}$. Secondary ionization events enhance the overall ionization in a dense environment.
The x-ray pulse with 10\,fs FWHM temporal profile is centered at $t=0$\,fs.
}
\label{fig:time-evolution}
\end{figure*}

The coherent scattering patterns depend on the presence of the atomic bound electrons as well as on the atomic positions.
The XMDYN and XATOM simulations allow to analyze their change due to radiation damage for both diamond and for the isolated carbon atom cases. 
Radiation damage is initiated by atomic photoionization events. 
In case of isolated carbon atoms Auger decays contribute approximately to the same extent to the overall ionization. 
At the maximum fluence in our study, $\sim 35$\,\% of the atoms are photoionized (Fig.~\ref{fig:time-evolution}.a). 
Although the absorbed energy is 10\,keV per photon, almost all of this energy is taken away from the atom by the high energy photoelectron.
The picture is different when the atom is embedded in a crystal environment (Fig.~\ref{fig:time-evolution}.c).
The high energy photoelectrons stay within the medium and distribute their energy by causing further ionization via secondary ionization events. 
As a consequence, neutral atoms disappear early in the pulse and by the end even fully stripped carbon ions ($\rm{C^{6+}}$) appear. 
Many electrons are promoted to (quasi-)free states within the sample.
This also illustrates the importance of secondary ionizations in the progress of radiation damage in a dense environment~\cite{vinko2015,vinko2012,beata2005}.
In the center of focus the sample absorbs 3.5~keV energy per atom that heats up the plasma electrons beside the ionizations. 
Despite the high charge states recombination remains negligible during the pulse (number of events less than 1~\% per atom in the simulation) due to the extreme conditions. 
%when no (or at least highly reduced) effect of the photoelectron 

%For the flattop spatial beam every super-cell experiences 
%uniform fluence. As a consequence of no particle transport from one super-cell to another, the highly energetic 
%photoelectrons always remains confined inside the super-cell. Due to this confinement the degree of ionization 
%increases for the flattop fluence case therefore, the degree of ionization damage is over-estimated. 
%As the FWHM of the temporal pulse envelope is 10\,fs therefore, the number of recombination
%events is less then 1\% of the carbon atoms at these time scales. For the spatial Gaussian beam the ionization damage from the 
%max. fluence super-cells is over-estimated. Whereas, the damage from the lower fluence super-cells which resides on 
%the wings of the Gaussian beam is under-estimated because no photoelectrons from high fluence region can interact with those regions. 
%The strength of the scattering signal depends on the central region of the spatial beam therefore, in our study the extent of damage 
%due to ionization is over-estimated in that region for the Gaussian case as well.

\begin{figure}

\includegraphics[width=7.0cm,height=6.0cm]{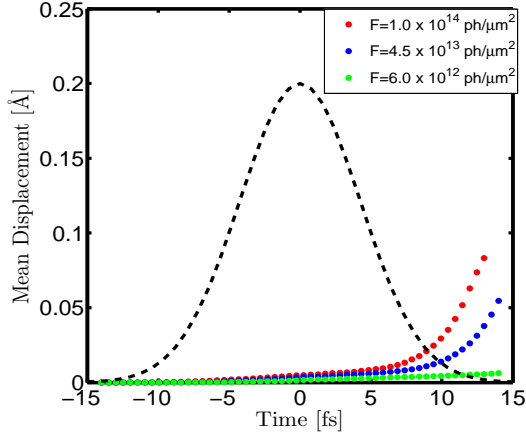}
\caption{Mean displacement of the atoms for fluences $\mathcal{F}_\text{high}\,=\,$1$\times10^{14}\,\mu m^{-2}$ (red dots),
 $\mathcal{F}_\text{mid}\,=\,$4.5$\times10^{13}\,\mu m^{-2}$ (blue dots) and $\mathcal{F}_\text{low}\,=\,$6.0$\times10^{12}\,\mu m^{-2}$ (green dots). The gaussian temporal 
pulse envelope is also depicted with the dashed black line. $\mathcal{F}_\text{high}$ is the fluence for the flattop profile, which is also the maximum fluence in the present study. 
$\mathcal{F}_\text{mid}$ and $\mathcal{F}_\text{low}$ are two values representing intermediate and low fluences taken from the gaussian 
profile case. The mean atomic displacement remains below the achievable resolution ($\sim 1.2 \rm{\AA}$) at 10keV for all the cases.}
\label{fig:mean-disp}
\end{figure}

Figure~\ref{fig:mean-disp} represents the mean displacement of the carbon atoms during the pulse. 
The average atomic displacement is much below the maximum achievable resolution, $\sim 1.2\rm{\AA}$ at 10\,keV, even at the highest fluence.
This suggests that the patterns are affected predominantly due to the bound-electron loss through the modification of atomic scattering form factors.
Despite the heavy ionization, atomic displacements remain negligible during the ultrashort pulse duration due to the highly symmetrical sample environment. 
We note here again that in our calculations we neglected the chemical bonds. In low fluence regions bonds may survive and stabilize the structure against the emerging Coulomb forces. 
As the observed displacements are far below the resolution even without any stabilization due to bonds, bondless modeling of the current scenario is applicable.

%It can be expected that neglecting the bonds can lead to larger kinetic energies but it can be seen that the average displacements are negligible in our study. 
%The reason for the average displacements to be very less is that for the considered fluences ionization is so fast that the atoms have
%no chance to move before bonds are broken. Therefore bondless modeling yields similar ion yields and kinetic energies. 

%It can also be seen from the Fig.~\ref{fig:mean-disp} that the average displacements increases as a function of time which is an indication of the fact that 
%screening of the charges does not play an important role at such short time scales.

\emph{Effect of the PBC approach on the dynamics.}
While ionic motion is negligible during the pulse, fast photoelectrons can travel long distances. 
However, PBC confines all plasma electrons artificially within the supercell they have been created in.
Neglecting particle transport may lead to error in ($i$) local plasma electron density and ($ii$) local energy density.
Whenever a photoelectron is ejected it leaves behind a positive charge located on an ion.
If we consider Coulomb interaction only, a positive space charge would build up in a central cylinder because of  photoelectron escape.
Photoelectron trapping within the interaction volume would start early in the pulse, at an average ion charge as low as +0.005.
An analogous phenomenon was discussed for finite samples in the literature~\cite{haureige2012}.
However, photoelectrons cause secondary ionization as well, so an atomic bound electron is promoted to a low energy continuum state.
If this slow electron was created in an outer region, it can efficiently contribute to the screening of the space charge the photoelectron left behind.
Based on these arguments we can conclude that ($i$) considering the interaction region to be neutral is a good approximation and ($ii$) in all regions we overestimate 
the energy density by confining fast photoelectrons within a supercell. Similarly, as the Coulomb forces are the driving forces of the ionic motions, we may also overestimate 
the atomic/ionic displacements. In our study eventually the effect on the scattering signal is relevant, as will be discussed in the next section.

\subsection{\label{sec:Scattering} Scattering with damage}

\begin{figure*}
\includegraphics[width=12.2cm]{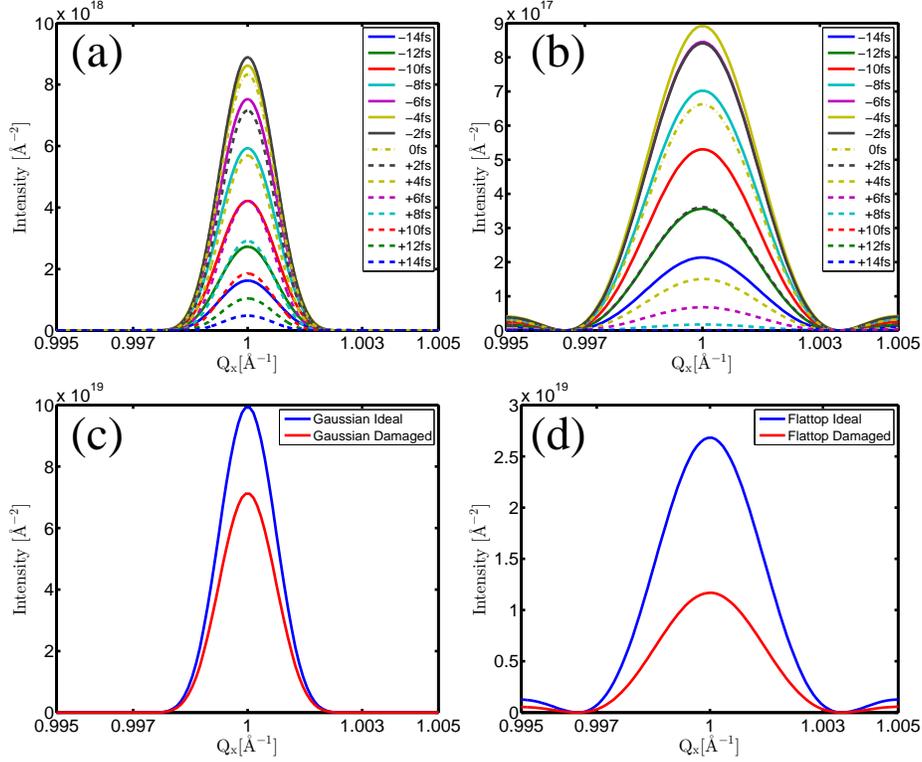}
\caption{Snapshots of the scattering intensity for reflection (1 1 1) along the $\rm{Q}_y = \rm{Q}_z = 1\,\AA^{-1}$ line in reciprocal space: (a) 
gaussian spatial beam profile, (b) flattop spatial beam profile. Solid and dashed lines with the same color correspond to the same instantaneous irradiating x-ray intensities. 
Note that the negative and the corresponding positive times are of equal intensity during the rise and fall of the pulse envelope.
(c,d) Total time integrated scattering signal for gaussian and flattop spatial beam profiles, respectively. Note the different vertical axis scales.}
\label{fig:intensity-evol}
\end{figure*}

In this section we analyze the changes of the Bragg peak intensity profiles in reciprocal space due to the severe radiation damage. 
In Fig~\ref{fig:intensity-evol}(a) and~\ref{fig:intensity-evol}(b) snapshots of the 1D Bragg peak profiles in reciprocal space are depicted for the reflection $Q$\,=\,(1 1 1) for Gaussian and flattop spatial 
beam profiles, respectively. Two apparent features can be seen, valid for other reflections as well.\\ 
($i$) The width of the Bragg peak does not change during the pulse. 
This is consistent with the expectation based on the negligible ion displacements: no visible Debye-Waller-like broadening occurs. 
However, the widths are different for the gaussian and flattop cases. The reason is the difference between the size of the illuminated parts of the crystal. 
In the flattop profile case the focus size defines strictly the region exposed. 
On the other hand a gaussian profile has no sharp edge and therefore illuminate a larger region, yielding a narrower Bragg peak and a larger effective crystal size.\\
($ii$) Snapshots of the Bragg peak intensities behave differently for flattop and gaussian beams.
The snapshots of the Bragg intensities depend not only on the scattering power of the sample but also on the instantaneous x-ray intensity. 
However, as the instantaneous x-ray intensities are equal at the same time before and after the maximum of the pulse, a direct comparison of the corresponding 
snapshots of the Bragg profiles reflects exclusively the effect of different damage extents.
In the gaussian profile case these corresponding curves show small difference only, indicating that a significant contribution is coming from regions in the crystal 
suffering little damage (Fig.~\ref{fig:intensity-evol}.a).
In contrast, applying a flattop pulse profile, the scattering pattern is formed only from extensively ionized parts of the crystal.
A consequence of the loss of atomic bound-electrons is the decrease of the atomic form factors yielding significant signal drop for longer times (Fig.~\ref{fig:intensity-evol}.b).

\begin{figure*}
\includegraphics[width=12.2cm]{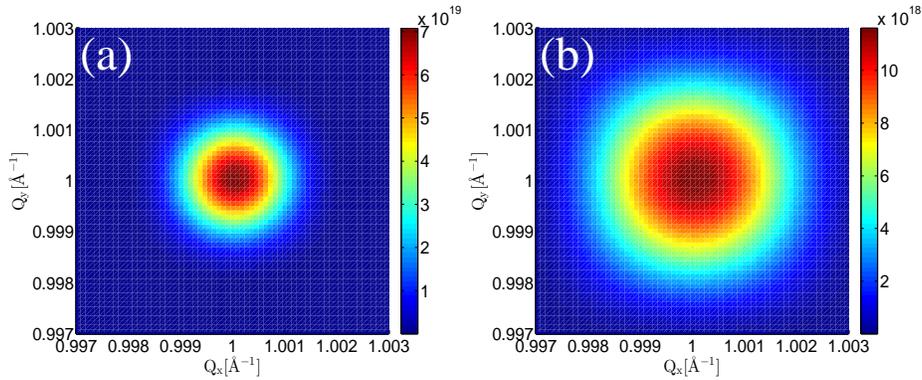}
\caption{ Contour plot for the Bragg spot of reflection (1 1 1) in the $\rm{Q}_z = 1\,\AA^{-1}$ plane in reciprocal space: 
(a) Gaussian beam profile; (b) flattop beam profile.}
\label{fig:2d-cuts}	
\end{figure*}

The above findings are reflected by the time integrated signals that correspond to the situation one would encounter in an experiment 
(1D cut:Fig.~\ref{fig:intensity-evol}.c,d; 2D cut: Fig.~\ref{fig:2d-cuts}). Note that for the gaussian spatial profile there is only a 
small decrease of the signal compared to the ideal (no damage) case. 

\emph{Effect of the PBC approach on the x-ray scattering patterns.}
In section \ref{sec:Damage} we discussed that the PBC approximation overestimates ionization and atomic displacements, and therefore radiation damage throughout the sample.
It means that the method gives an upper bound to the effect of radiation damage on the scattering patterns. 
A trivial lower bound is the case without any radiation damage.

%For a gaussian spatial profile the scattering signal is formed predominantly by contribution from the central part as the gaussian introduces where applying PBC overestimates radiation damage. 
%Therefore we conclude that the approach gives an upper bound to the radiation damage effects on the scattering patterns.

\section{\label{sec:level1}Conclusions}

We presented a methodology for the simulation of x-ray scattering patterns from serial femtosecond crystallography experiments with a high intensity x-ray beam. 
Our approach includes the simulation of radiation damage within the sample with the codes XMDYN and XATOM as well as the calculation of the patterns using the code XSINC.
In the approach the crystal is divided into smaller units. 
The time evolution (the radiation damage process) of these units is calculated using periodic boundary conditions. 
Finally a nanocrystal is assembled from the small units for the calculation of the patterns integrated over the pulse.

As a demonstration we investigated spatial pulse profile effects on the Bragg peaks for a diamond nanocrystal. 
We found that if a gaussian profile is used (assuming realistic XFEL parameters such as tight focus and ultrashort pulse duration), the time integrated signal 
intensity is reduced only by a small amount compared to the damage-free case.
For a flattop profile the decrease is much more significant. The intensity reduction is due to the change of the form factors caused by ionizations.
In both cases the width of the Bragg peak was connected to the size of the illuminated region in the crystal, but was not affected by damage.
We analyzed the shortcoming of the periodic boundary condition approach. The method overestimates radiation damage in the interaction region, so 
it gives an upper bound to the effect of radiation damage on the patterns. In the future the simulation method developed here is to be applied to more complex scenarios.

%The only shortcoming of the approach is the exclusion of transport of highly energetic electrons from high fluence region to the low fluence region.
%From the results it can be concluded 
%that the extent of ionization has been over-estimated in our calculation as the highly energetic electrons never escaped the super-cells and lead to more ionization
%events due to impact ionization. In the future the simulation method developed here is to be applied to more complex scenarios.

\section*{Acknowledgement}
This work has been supported by the excellence cluster ``The Hamburg Centre for Ultrafast Imaging (CUI): Structure, Dynamics and Control of 
Matter at the Atomic Scale'' of the Deutsche Forschungsgemeinschaft.

\bibliography{xsinc-methodolgy}% Produces the bibliography via BibTeX.

\listoffigures
\end{document}